# Quantum dot self-assembly driven by a surfactant-induced morphological instability

*Ryan B. Lewis,[1,]\* Pierre Corfdir,[1] Hong Li,[1,2] Jesús Herranz,[1] Carsten Pfüller,[1] Oliver Brandt[1] and Lutz Geelhaar[1]*

[1] Paul-Drude-Institut für Festkörperelektronik, Hausvogteiplatz 5-7, 10117 Berlin, Germany

[2] Institut für Physik und IRIS Adlershof, Humboldt-Universität zu Berlin, Zum Großeb Windkanal 6, 12489 Berlin, Germany

*Email: lewis@pdi-berlin.de

In strained heteroepitaxy, two-dimensional (2D) layers can exhibit a critical thickness at which three-dimensional (3D) islands self-assemble, relieving misfit strain at the cost of an increased surface area. Here we show that such a morphological phase transition can be induced on-demand using surfactants. We explore Bi as a surfactant in the growth of InAs on GaAs(110), and find that the presence of surface Bi induces Stranski−Krastanov growth of 3D islands, while growth without Bi always favors 2D layer formation. Exposing a static two monolayer thick InAs layer to Bi rapidly transforms the layer into 3D islands. Density functional theory calculations reveal that Bi reduces the energetic cost of 3D island formation by modifying surface energies. These 3D nanostructures behave as optically active quantum dots. This work illustrates how surfactants can enable quantum dot self-assembly where it otherwise would not occur.

Three-dimensional (3D) islands that act as quantum dots (QDs) are typically synthesized by a self-assembly process known as Stranski−Krastanov (SK) growth, one of the three fundamental modes of epitaxy.[1,2] In general, the preferred growth mode of a system is determined by energetic considerations (surface, interface, strain), which are fixed by the choice of adsorbate and substrate. To some extent these thermodynamic constrains can be overcome kinetically, either by adjusting deposition conditions[3] or by changing the surface chemistry using surface segregating elements,[4–8] allowing 3D island formation to be kinetically suppressed in favor of two-dimensional (2D) Frank−van der Merwe (FM) growth. However, control over the energetically preferred growth mode, as well as inducing 3D island formation when 2D growth is favored, has remained elusive. And as a result, QD synthesis has been restricted in terms of materials and substrate orientations.

The (In,Ga)As/GaAs(100) system is the most extensively explored SK growth system, demonstrating (In,Ga)As QDs suitable for quantum optics applications requiring single photon emission[9,10] and emission of entangled photon pairs.[11,12] While (In,Ga)As favors SK growth for a wide range of compositions and deposition conditions on GaAs(100), on other low-index GaAs surfaces such as (110) and (111), growth occurs via a 2D FM mode and misfit strain relaxes plastically.[13–16] Due to the low surface energy of GaAs(110), {110} facets are often present in self-assembled GaAs nanostructures such as nanowires and the growth of QDs on these structures is of interest for high brightness single photon sources.[17,18]

Here we show that surfactants can provoke morphological phase transitions in strained layers, inducing the formation of 3D islands "on-demand". We investigate Bi as a surfactant in the growth of InAs on GaAs(110) by molecular beam epitaxy (MBE). The presence of surface Bi is found to alter the fundamental growth mode of InAs, inducing 3D island formation by the SK mechanism. Furthermore, surface Bi can provoke a morphological phase transition in static 2D InAs layers, inducing a rapid rearrangement of the 2D layer into 3D islands. Density functional theory (DFT) calculations reveal that surface Bi reduces the energetic cost of 3D island formation by altering the surface energy of the GaAs and InAs surfaces. Photoluminescence (PL) spectroscopy on these novel nanostructures demonstrates that they behave as optically active QDs. This work illustrates how modifying surface energies with surfactants can allow for unprecedented control over nanostructure self-assembly.

Samples were grown by MBE on 2″ GaAs(110) wafers. Fluxes were provided by effusion cells for Ga, In and Bi, and a valved cracker for $As_2$. A substrate temperature of 420°C and an As flux of 1.1 nm/s GaAs equivalent growth rate were maintained for the entire deposition process. GaAs buffer layers (50−150 nm thick) and capping layers (50 nm thick) were deposited at 0.28 nm/s. For experiments exploring concurrent In and Bi deposition, the Bi flux was initiated 20 s before the In flux to produce a stable Bi coverage. The Bi beam equivalent pressure (BEP) was $2\times10^{-6}$ mbar and the In flux was 0.1 ML/s. It is expected that Bi will not alloy with InAs at these growth conditions, due to the large As/In flux ratio.[19] In further experiments, Bi was deposited subsequent to InAs, and in this case the Bi flux was maintained for 30 s before cooling the sample at 2°C/s while maintaining the As$_2$ flux, until the substrate was below 350°C. Samples for photoluminescence (PL) studies were capped with 50 nm of GaAs.

Density functional theory calculations were performed with the Vienna Ab Initio Simulation Package (VASP).[20,21] The *d* electrons for Ga, In, and Bi were considered as valence states, and the (110) surfaces were modeled with a slab consisting of nine atomic layers and a vacuum of 20 Å. Pseudohydrogen atoms with partial charge of 0.75 (1.25) were used to passivate the dangling bonds of the bottom surface As (Ga) atoms.[22] The local-density approximation (LDA) was chosen for the exchange-correlation functional. The calculated surface energies were 50.6 meV/Å$^2$ and 40.2 meV/Å$^2$ for GaAs (110) and InAs (110), respectively, in good agreement with other calculations.[23]

Magneto-PL experiments were performed at 4.2 K in a confocal setup operating in Faraday geometry (magnetic field parallel to the [110] direction) with magnetic field strength between 0 and 8 T. The samples were excited using a Ti:Sapphire laser emitting at 790 nm, and the laser spot diameter and power at the surface of the samples were 1 μm and 10 μW, respectively. The PL signal was dispersed using a monochromator and detected with a charge-coupled device camera.

Figure 1 shows atomic force microscopy (AFM) topographs after deposition of 2.1 monolayers (MLs) of InAs on GaAs(110), without and with the presence of a Bi flux during the InAs deposition. We note that 2.1 MLs on GaAs(110) corresponds to 1.5 MLs on GaAs(100). In the absence of Bi [Fig. 1(a)], InAs forms a 2D layer with atomic terraces on the surface. The addition of a Bi flux [Fig. 1(b)] results in a drastically different surface morphology, showing an array of 3D islands with a density of 6×10$^9$ cm$^{-2}$ and an average island height of 4.3 nm (standard deviation 0.5 nm). The presence of 3D islands is in striking contrast to the expectation that InAs deposition on GaAs(110) always proceeds by a 2D FM mode. However, we note that the growth of InAs 3D islands has been demonstrated on GaAs(110) substrates covered by thin (In,Ga)As[24,25] and AlAs layers.[26] The islands in Fig. 1(b) are elongated in the [1$\bar{1}$0] direction, which may be a result of the elastic anisotropy, as InAs is less stiff in the [1$\bar{1}$0] direction than in the perpendicular [001] direction, and elongation along [1$\bar{1}$0] preferentially relaxes [001] strain. Alternatively, the elongation could result from higher adatom diffusion along [1$\bar{1}$0]. During the growth of this sample, the reflection high energy electron diffraction (RHEED) pattern showed a streaky-to-spotty (2D-to-3D) transition at a thickness of about 1.9 ML. The observation of a critical thickness for 3D island nucleation is consistent with SK growth. Conversely, increasing the InAs thicknesses to 4.2 MLs in the absence of Bi does not result in 3D nanostructure formation (see Supplemental Material).

Energy-dispersive x-ray spectroscopy investigation of samples grown with Bi BEPs higher than 2×10$^{-6}$ mbar only indicated the presence of Bi in the InAs 3D islands at a Bi BEP of $1\times10^5$ mbar. We note that Bi is not expected to form a compound with InAs at the high As/In flux ratios used in this study,[19] and the incorporation of small amounts of Bi is not expected to degrade the optoelectronic properties of InAs. In fact, the presence of surface Bi during InAs QD growth on GaAs(100) has been shown to greatly improve QD photoluminescence.[27]

The above results show that the presence of surface Bi can alter the fundamental growth mode of InAs on GaAs(110) from a 2D FM mode to a 3D mode. It is expected that 3D islands never form in the absence of Bi because the critical thickness for 3D island formation is larger than the critical thickness for plastic relaxation (previously reported to be 2–3 MLs).[15] With the presence of surface Bi, the situation is reversed. To further explore the effect of surface Bi on the morphological stability of InAs, we expose a 2.1 ML thick 2D InAs layer grown without Bi [Fig. 1(a)] to a subsequent Bi flux for 30 s. Figure 2 (a) shows the resulting surface topography, which is compared to 3D islands grown at the same conditions but with simultaneous InAs and Bi deposition [Fig. 2(b), same sample as in Fig. 1(b)]. Remarkably, exposing the 2D InAs layer to Bi results in a rearrangement of the InAs into 3D islands. Compared to the sample grown with codeposition of Bi and InAs [Fig. 2(b)], the island density is about 3 times higher and the dots are more symmetric, suggesting that island nucleation occurs more rapidly. The initiation of the Bi exposure produces an abrupt transition of the RHEED pattern from 2D streaks to 3D spots. Therefore, surface Bi can provoke a morphological phase transition in a static InAs layer, indicating that the effect of surface Bi goes beyond modifying adatom kinetics during InAs growth. In the absence of Bi, the 2.1 ML InAs thickness is below the critical thicknesses for both 3D island formation and plastic relaxation, but in the presence of Bi the layer is thicker than the critical thickness for 3D island formation. Therefore, the morphological stability of the layer can be controlled externally through the Bi coverage, allowing a morphological phase transition to be induced on-demand. This unprecedented external control opens up new possibilities for 3D nanostructure self-assembly.

Surface segregating elements such as Sb, Te and Bi, have been previously investigated in the growth of (In,Ga)As, although only on the (100) surface of GaAs. In contrast to the present findings for Bi, Sb and Te have been shown to inhibit 3D growth, by dramatically reducing adatom diffusion.[6,28] While Bi has been found to actually increase adatom diffusion during 2D growth,[29] the surfactant effect of Bi in III-V epitaxy has been only sparsely explored.[27,29–33]

The formation of 3D islands during epitaxy has been investigated theoretically in terms of the Asaro-Tiller-Grinfeld (ATG) linear instability theory,[34,35] as well as nucleation theory.[36] The roughening of a 2D layer is driven by strain relaxation but inhibited by surface energy. To elucidate the effect of surface Bi on the surface energy, we carried out DFT calculations. We consider various surface configurations and compare the surface formation energies relative to the energy of the relaxed bulk-truncated 1×1 GaAs(110) surface. The relative surface formation energy is defined as

$$\Delta \gamma = [E_{surf} - E_{ref} - \sum_i n_i \mu_i]/A \qquad (1)$$

where $E_{surf}$ and $E_{ref}$ are the total energies calculated for the surface structure and the reference 1×1 GaAs(110) structure, $n_i$ and $\mu_i$ are the excess number of atoms (compared to the reference structure) and the chemical potential of element *i*, respectively, and *A* is the surface area. The chemical potentials of GaAs, InAs, and Bi were deduced from the calculated total energies for their bulk structures. The relative surface energy includes contributions from the strain energy, as well as energy changes due to the surface and interfaces. In the absence of InAs, we find that the most stable configuration for GaAs(110) is a single Bi adlayer covering the 1×1 GaAs surface. Similarly, for InAs on 1×1 GaAs(110) the 1×1 InAs surface structure was found to be the most stable (see Supplemental Material). We note that experimentally we observe only a 1×1 RHEED pattern and this is consistent with previous observations of InAs deposition on GaAs(110).[15] Therefore, for calculations involving InAs epilayers on GaAs(110) we consider only 1×1 surfaces and a single Bi adlayer. We note that growth of InAs on top of a Bi monolayer is highly energetically unfavorable, resulting in large positive Δγ values (see Supplemental Material). Therefore, during InAs growth the Bi layer is expected to remain on the surface.

Figure 3(a) displays the relative surface energy as a function of the InAs wetting layer thickness on GaAs(110) with and without a Bi adlayer. In the absence of Bi, Δγ is negative for an InAs coverage of 1 ML due to the lower InAs surface energy, but positive for higher coverages as a result of strain. Therefore, it is energetically preferable for InAs to wet the GaAs(110) surface. However, beyond 1 ML coverage, the layer is metastable. Since Δγ increases with increasing InAs coverage, eventually a critical thickness is reached where the layer relaxes plastically (reported to be 2–3 ML[15]). With the addition of a Bi adlayer, the situation is strikingly different. In this case, the Bi terminated GaAs(110) surface (0 ML InAs) shows a large negative Δγ of –7.4 meV/Å$^2$. Thus, Bi behaves as a surfactant on GaAs, reducing the surface energy. With increasing InAs wetting layer thickness Δγ increases monotonically, due to the increasing strain energy. In contrast to the case without Bi, the lowest energy surface does not contain an InAs wetting layer. For InAs coverages of 2 and 3 ML, the presence of the Bi adlayer increases Δγ. This can be understood by considering the surface structures presented in Fig. 3 (b), for 0–2 ML InAs coverage with and without Bi, where stronger bond relaxation and re-hybridization are found to occur on the bare GaAs and InAs surfaces. As Bi suppresses the distortion of the InAs lattice, it is expected that the InAs strain energy is higher in the presence of Bi. This could explain why the relative surface energy is larger with Bi than without at higher InAs thicknesses.

The DFT results indicate that the Bi adlayer reduces the relative surface energies of bulk GaAs(110) and InAs(110) (see Supplemental Material for bulk InAs). Furthermore, surface Bi reduces the thermodynamic driving force for InAs wetting on GaAs(110), as the lowest energy state is Bi-terminated GaAs(110). In other words, 2D InAs layers under the Bi adlayer always have an energy greater than Bi-terminated GaAs(110) plus bulk InAs, suggesting that the Volmer-Weber growth mode may be possible on the Bi-terminated surface. Finally, Bi inhibits bond relaxation at the InAs surface, increasing the InAs strain energy. All of these effects will favor 3D island formation compared to the case without Bi. For SK growth, the energy barrier for 3D island formation is believed to control the critical thickness for islanding.[2,36] This energy barrier is related to the surface energy as well as strain energy. Therefore, the DFT results are consistent with Bi reducing the energy barrier for 3D island formation. Such a surfactant-induced lowering of the energy barrier for 3D island formation has been predicted theoretically.[37]

To investigate whether Bi surfactant-induced 3D islands can behave as optically active QDs, a sample with about 1.5 ML of InAs grown with In and Bi codeposition and then capped with GaAs was investigated by PL. The insets of Fig. 4(b) presents AFM topographs from an uncapped sample with a similar InAs nominal thickness of 1.7 MLs (below the 2D-3D RHEED transition), showing small 3D islands of about 10 nm in diameter on the surface at a density of about $10^8$ cm$^{-2}$. The capped sample is expected to exhibit similar islands, and we speculate that with further InAs deposition these small islands enlarge into the islands presented in Figs. 1 and 2. A PL spectrum taken on the capped sample is displayed in Fig. 4(a), where a set of transitions with linewidths of about 0.36 meV (resolution limited) is observed between 1.29 and 1.34 eV. For samples grown without Bi these transitions are not detected, suggesting that they originate from the surfactant-induced 3D islands. An enlarged PL spectrum of a typical transition is shown in the inset. With increasing excitation power, the intensity of this line increases linearly (see Supplemental Material), showing that it is related to the recombination of a neutral exciton. The emission energy range in Fig. 4(a) corresponds to the one usually observed for excitons strongly confined in SK InAs QDs grown on GaAs(100). To obtain quantitative information on the size of the 3D islands giving rise to these transitions, we followed their energy in a magnetic field. Figure 4(b) shows the evolution between 0 and 8 T of the transition shown in the inset of Fig. 4(a). This line splits into two transitions, whose energies follow a parabolic dependence with increasing magnetic field. A parabolic fit to these emission energies[38] yields an exciton g-factor $g_1 = 2.8$ and a diamagnetic shift $g_2 = 7.5$ µeV/T$^2$. For neutral excitons, $g_2$ is proportional to the electron-hole coherence length ($L_{eh}$) in the plane perpendicular to the magnetic field.[38] Measuring $g_2$ for a total of 14 transitions, we obtain an average in-plane $L_{eh}$ of 5.4 ± 1.2 nm. This value, which corresponds to the radius of the 3D islands giving rise to light emission in the 1.3 to 1.4 eV range, is consistent with the size of the islands shown in the insets of Fig. 4(b). We therefore attribute the observed PL emission to these objects. These results demonstrate that the surfactant Bi can enable the growth of optically active QDs on a surface where QD growth is otherwise inhibited.

In conclusion, morphological phase transitions in strained films can be induced by externally modifying surface energies with surfactants. We have used this approach to synthesize optically active quantum dots where they would otherwise not form. This unprecedented external control over the self-assembly of 3D nanostructures paves the way to realizing quantum dots with new materials and on new substrates, and provides an experimental framework to test theories of strained heteroepitaxy.

Acknowledgments

R.B.L. acknowledges funding from the Alexander von Humboldt Foundation. P.C acknowledges funding from the Fonds National Suisse de la Recherche Scientifique through Project No. 161032. The authors are grateful to M. Höricke and C. Stemmler for MBE maintenance, M. Ramsteiner and G. Paris for help with the magneto-PL setup, and V.M. Kaganer for a critical reading of the manuscript.

FIGURE CAPTIONS

FIG. 1. Surface topographs after deposition of 2.1 ML of InAs on GaAs(110) (a) in the absence of Bi and (b) with a Bi flux. In the absence of Bi the surface is atomically smooth, while growth with the Bi flux results in 3D islands. The scale bar and crystallographic directions shown in (a) also apply to (b).

FIG. 2. Comparison of InAs 3D islands formed by two approaches: (a) exposing a static 2.1 ML 2D InAs layer to Bi and (b) depositing 2.1 MLs of InAs while simultaneously depositing Bi to induce SK growth. The height and density of the 3D islands in (a) are 4.6 ± 1.3 nm and about $1.7 \times 10^{10}$ cm$^{-2}$, respectively. The scale is the same for both images and the [1$\bar{1}$0] direction is approximately upward. The total volume of the 3D islands in (a) and (b) corresponds to approximately 1.9 and 1.7 ML of 2D growth, respectively, assuming the islands are composed of coherent InAs.

FIG. 3. Relative surface energy and structure of InAs/GaAs(110) for varying InAs coverage with and without a Bi adlayer. (a) Relative surface formation energies calculated by DFT for the Bi terminated and bare surfaces assuming the 1×1 epitaxial continued layer structure. The in-plane lattice parameter for the entire structure is fixed to that of bulk GaAs. (b) Surface structures for 0–2 ML InAs coverage without (upper row) and with (lower row) the Bi adlayer. Bi suppresses the bond relaxation of the GaAs and InAs surfaces.

FIG. 4. (a) Photoluminescence spectrum from InAs QDs on GaAs(110) displaying a series of narrow transitions. The inset shows emission from a single InAs QD with full width at half maximum 0.36 meV (resolution limit). (b) Magneto-PL intensity map from the QD transition shown in the inset in (a). The quadratic fits to the QD emission energy as a function of magnetic field account for Zeeman and diamagnetic effects, yielding an exciton g-factor $g_1 = 2.8$ and a diamagnetic shift $g_2 = 7.5$ μeV/T$^2$. The insets show exemplary AFM images of small QDs from an uncapped sample grown with 1.7 ML InAs.

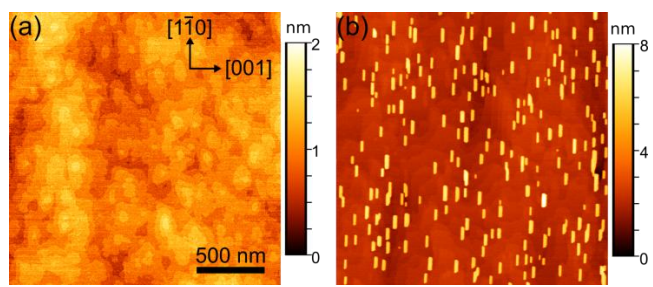

FIG. 1.

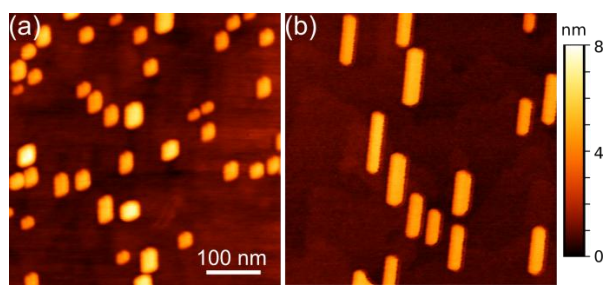

FIG. 2.

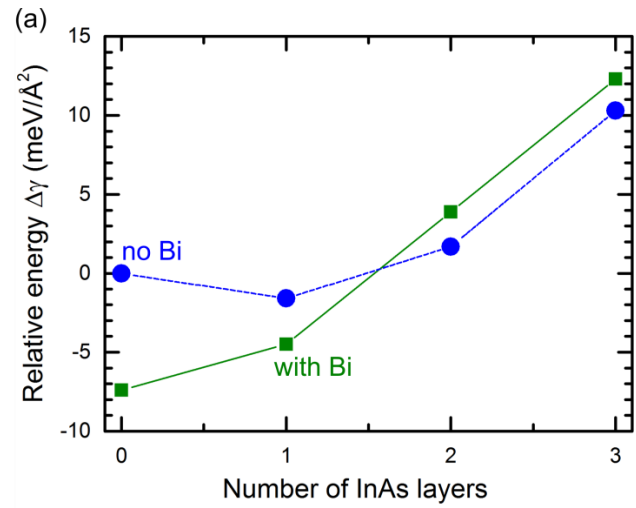

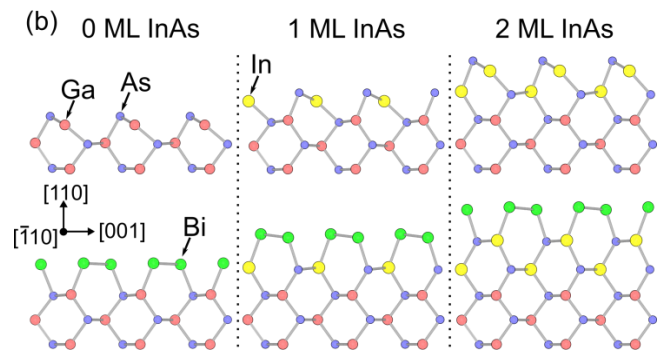

FIG. 3.

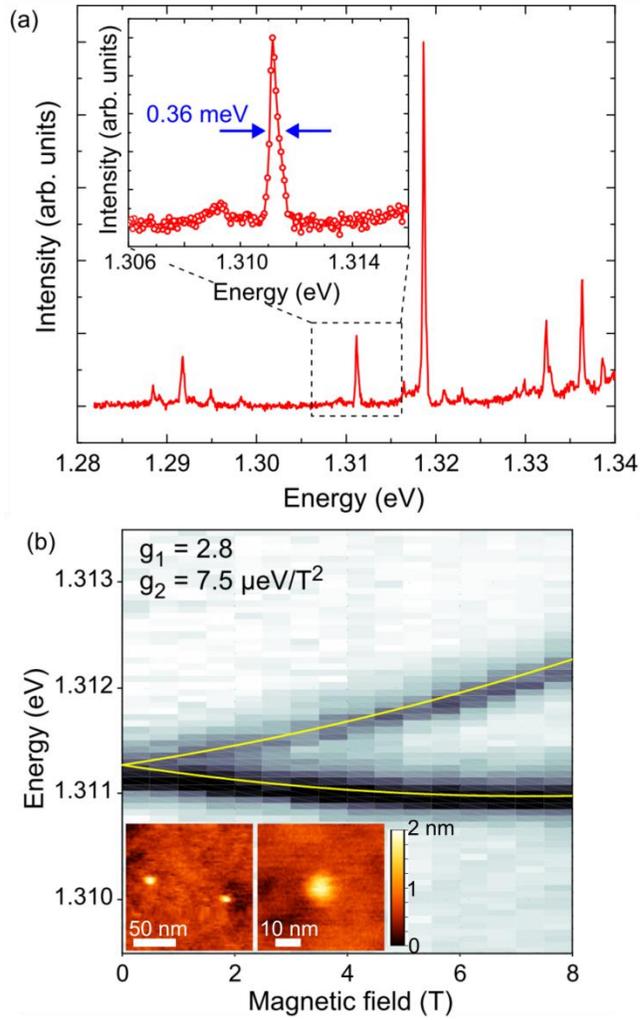

FIG. 4.

# SUPPLEMENTAL MATERIAL

# Quantum dot self-assembly driven by a surfactant-induced morphological instability

*Ryan B. Lewis,[1,]\* Pierre Corfdir,[1] Hong Li,[1,2] Jesús Herranz,[1] Carsten Pfüller,[1] Oliver Brandt[1] and Lutz Geelhaar[1]*

[1] Paul-Drude-Institut für Festkörperelektronik, Hausvogteiplatz 5-7, 10117 Berlin, Germany

[2] Institut für Physik und IRIS Adlershof, Humboldt-Universität zu Berlin, Zum Großeb Windkanal 6, 12489 Berlin, Germany

*Email: lewis@pdi-berlin.de

## CONTENTS

I. ATOMIC FORCE MICROSCOPY IMAGE OF A 4.2 ML InAs LAYER
II. SURFACE ENERGY CALCULATIONS FOR DIFFERENT CONFIGURATIONS
III. EXCITATION POWER DEPENDENCE OF THE QD PHOTOLUMINESCENCE

## I. ATOMIC FORCE MICROSCOPY IMAGE OF A 4.2 ML InAs LAYER

Figure S1 displays an atomic force microscopy (AFM) topograph of a 4.2 monolayer (ML) InAs layer grown on GaAs(110) in the absence of Bi. The surface consists of large islands that are elongated along [1$\bar{1}$0] with heights up to 6 nm. These islands show small ridges running along [1$\bar{1}$0]. This morphology is consistent with scanning tunneling microscopy and transmission electron microscopy micrographs of InAs layers on GaAs(110) containing a high density of dislocations.[1]

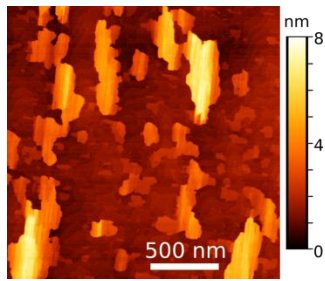

FIG. S1. AFM topography image of a 4.2 ML InAs layer grown on GaAs(110) without Bi. The [1$\bar{1}$0] direction is rotated clockwise by about 5° from the vertical direction.

## II. SURFACE ENERGY CALCULATIONS FOR DIFFERENT CONFIGURATIONS

For the density functional theory (DFT) calculations, the four bottom layers of the nine layer thick GaAs slab were fixed in their bulk geometry together with the pseudohydrogen atoms, while the other layers were allowed to relax until all forces were converged to less than 0.005 eV/Å$^2$. A dipole correction[2] was applied to accelerate the convergence of total energies with respect to the vacuum size. The local-density approximation (LDA) was chosen for the exchange-correlation functional and the plane wave cutoff was set to 400 eV. An 8×6×1 Monkhorst-Pack[3] k-point mesh was used for the 1×1 surface unit cell. Convergence of surface energy with respect to k-point sampling, energy cutoff, vacuum, and slab thickness is ensured within 2 meV/Å$^2$.

We consider the presence of Bi on GaAs (110) and InAs (110) surfaces assuming their bulk in-plane lattice constants. Two previously investigated surface unit cells are considered: the bulk-truncated 1×1 structure[4] and the 1×2 reconstructed missing-row structure.[5] We refer to an atom, one layer, two layers, and three layers of Bi on the 1×1 surface as a Bi adatom, adlayer, bilayer, and trilayer, respectively, while the adlayer* represents a Bi layer on the missing-row structure of the 1×2 cell. The corresponding Δγ for the relaxed structures (relative to the relaxed bulk-truncated 1×1 GaAs and InAs structures) are shown in Table SI. The negative Δγ for the Bi adlayer on both GaAs (110) and InAs (110) indicates that it is energetically favorable to cover these surfaces with Bi at thermodynamic equilibrium conditions. The 1×2 missing-row surface is less stable than the ideal 1×1 surface for GaAs, but more stable for InAs. In contrast, the Bi adatom, bilayer and trilayer structures show positive Δγ, indicating that they are unstable as their energy is higher than the sum of the bare surfaces and bulk Bi. The reasons for these higher energies are that there are more dangling bonds on the adatom surface, and that the strain energy increases in the Bi bilayer and trilayer compared to the Bi adlayer.

TABLE SI. The relative surface energies Δγ (in meV/Å$^2$) of Bi chemisorption on GaAs(110) and InAs(110) surfaces. The 1×1 surface was used for all calculations except for adlayer*, where the 1×2 missing-row surface was used.

|            | adatom | adlayer* | adlayer | bilayer | trilayer |
|------------|--------|----------|---------|---------|----------|
| GaAs(110)  | 19.4   | −7.1     | −7.4    | 23.0    | 26.0     |
| InAs(110)  | 18.0   | −8.0     | −5.3    | 25.4    | 21.6     |

Additionally, we consider Bi segregation on the GaAs(110) and InAs (110) 1×1 surfaces. To do this, the Bi position was varied from the 1$^{st}$ (the surface) to the 4$^{th}$ layer from the surface. The energy difference $\Delta E_i^{tot}$ between Bi on the $i^{th}$ layer $E_i$ and on the surface layer $E_1$ was calculated to determine the trend of Bi segregation.

$$\Delta E_i^{tot} = E_i - E_1 \tag{S1}$$

The resulting energy difference is plotted as a function of the Bi layer position in Fig. S2 for both the GaAs(110) and InAs(110) surfaces. In both cases, it is highly energetically unfavorable for Bi to reside below the surface, indicating a strong tendency for Bi surface segregation.

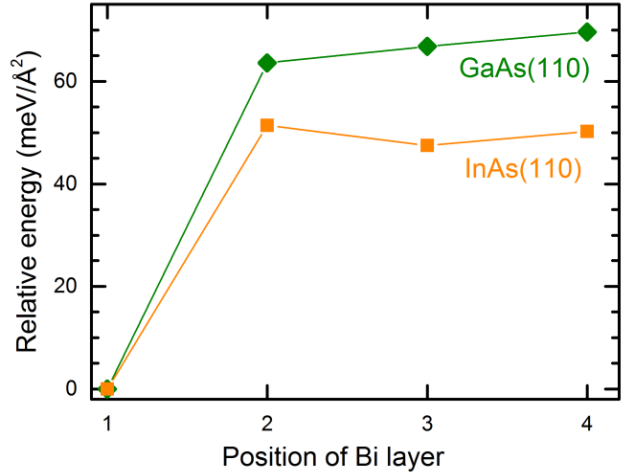

FIG. S2. The relative energy as a function of the Bi monolayer position within the slab. The energy is relative to the case when Bi is on the surface (position 1). Results for both GaAs(110) and InAs(110) 1×1 surfaces are shown.

The 1×2 missing-row surface was also considered for InAs layers on 1×1 GaAs(110) both with and without the presence of a Bi adlayer. Figure S3 shows the relative surface formation energies for these structures, along with those for the 1×1 surfaces that are shown in Fig. 3 in the main article. We note that the 1×2 missing-row InAs surface corresponds to 0.5 MLs of InAs. In the absence of Bi, $\Delta\gamma$ is much larger for the 1×2 surface than for the 1×1 InAs surface. Both 1×2 and 1×1 surface structures show similar energies when the Bi adlayer is included.

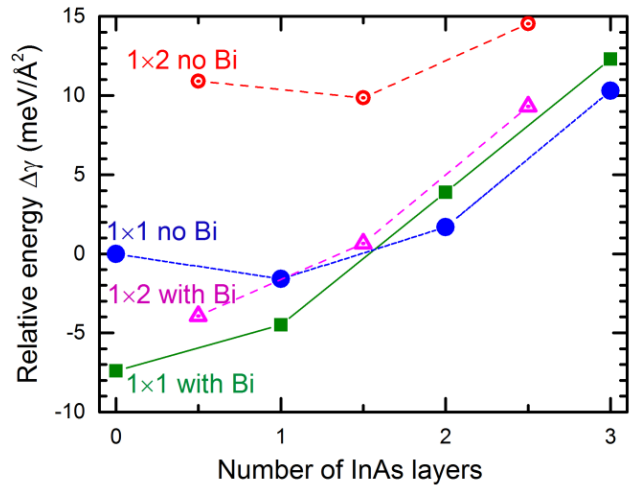

FIG. S3. Relative surface formation energies of InAs/GaAs(110) for varying InAs coverage with and without a Bi adlayer. The surface energies are shown for both the 2×1 missing-row and the 1×1 InAs surfaces. The 1×1 cell is assumed for the GaAs surface/interface.

### III. EXCITATION POWER DEPENDENCE OF THE QD PHOTOLUMINESCENCE

Figure S4 shows the excitation power dependence of the intensity of the photoluminescence line shown in the inset of Fig. 4(a) in the main article. The observed linear dependence indicates that this line corresponds to the recombination of a neutral exciton.

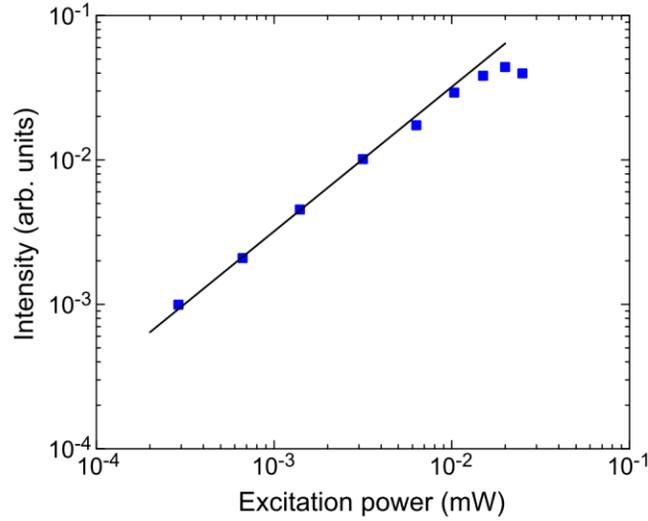

FIG S4. Photoluminescence intensity as a function of laser power for the emission line shown in the inset of Fig. 4(a). The solid line is a guide-to-the-eye highlighting the linear increase in intensity of this transition with increasing excitation power.